\documentclass[%
showpacs,preprintnumbers,
 amsmath,amssymb,
 aps,
 prd,
 longbibliography,
floatfix,
 lengthcheck,%
]{revtex4}

\usepackage{graphicx}
\usepackage{hyperref}

\newcommand{\ket}[1]{\ensuremath{|{#1}\rangle}}

\begin{document}

\title{The consequences of large $\theta_{13}$ for the turbulence signatures in supernova neutrinos}

\author{James P. Kneller}
\email{jpknelle@ncsu.edu}
\affiliation{Department of Physics, North Carolina State University, Raleigh, North Carolina 27695, USA}

\author{Alex W. Mauney}
\email{awmauney@ncsu.edu}
\affiliation{Department of Physics, North Carolina State University, Raleigh, North Carolina 27695, USA}

\begin{abstract}

The set of transition probabilities for a single neutrino emitted from a point source after passage through a turbulent supernova density profile have been
found to be random variates drawn from parent distributions whose properties depend upon the stage of the explosion, the neutrino energy and mixing
parameters, the observed channel, and the properties of the turbulence such as the amplitude $C_{\star}$. 
In this paper we examine the consequences of the recently measured mixing angle $\theta_{13}$ upon the neutrino flavor transformation in supernova when passing through turbulence.
We find the measurements of a relatively large value of $\theta_{13}$ means the neutrinos are relatively immune to small, $C_{\star}\lesssim 1\%$, amplitude turbulence but 
as $C_{\star}$ increases the turbulence effects grow rapidly and spread to all mixing channels. For $C_{\star} \gtrsim 10\%$ the turbulence effects in the high (H) density resonance mixing channels are 
independent of $\theta_{13}$ but non-resonant mixing channels are more sensitive to turbulence when $\theta_{13}$ is large .
\end{abstract}

\pacs{47.27.-i,14.60.Pq,97.60.Bw}
\date{\today}

\maketitle


\section{Introduction}

The progress in the field of supernova neutrinos over the past decade has frenetic. The rich phenomenology 
of neutrino collective effects \cite{Pantaleone:Gamma1292eq,Samuel:1993uw,Pastor:2001iu,Duan:2005cp,Hannestad:2006nj,Duan:2006an,Raffelt:2007cb,2008PhRvD..78l5015R,2011PhRvL.106i1101D,2011PhRvL.107o1101C,2011PhRvD..84h5023R,2012JPhG...39c5201G,2012PhRvL.108z1104C,2012PhRvD..85k3007S,2012PhRvL.108w1102M} (for a review see \cite{Duan:2009cd,Duan:2010bg}) have received the most attention but there has been an equally radical overhaul
of the Mikheyev, Smirnov \& Wolfenstein (MSW) \cite{M&S1986,Wolfenstein1977} effect as applied to supernova ever since it was realized by Schirato \& Fuller
\cite{Schirato:2002tg} that the shockwave racing through the stellar mantle could leave an imprint upon the neutrinos emitted from the
cooling proto-neutron star \cite{Schirato:2002tg,Takahashi:2002yj,Fogli:2003dw,Tomas:2004gr,Choubey:2006aq,Kneller:2007kg,Gava:2009pj}.
Recent studies indicate Earth matter effects may be minimal \cite{2012PhRvD..86h3004B}. 
From this ever-growing body of literature one now expects that the neutrino signal from the next supernova in our Galaxy will be pregnant with information. If
the signal can be decoded we might be able to both determine any unresolved properties of the neutrino and also to observe the explosion while it is still deep within the star.
Yet most, though not all, of these studies use spherically symmetric density profiles either in a parametrized form or taken from one-dimensional
hydrodynamical  simulations. While the use of one-dimensional hydrodynamical profiles for neutrino signal construction is probably adequate for certain
situations - such as neutrinos from Oxygen, Neon, Magnesium supernova \cite{2008PhRvD..78b3016L,2008PhRvL.100b1101D,2010PhRvD..82h5025C} which explode
in spherically symmetric simulations \cite{2006A&A...450..345K,2006ApJ...644.1063D,2010A&A...517A..80F} - it is now apparent that iron core collapse supernova should not be expected to be spherically symmetric. Large
scale inhomogeneities are created deep within the explosion and one observes turbulence during the neutrino heating/Standing Accretion Shock Instability phase \cite{2011ApJ...742...74M,2012arXiv1210.5241D,2012arXiv1210.6674O,2012ApJ...755..138H,2012ApJ...746..106P,2012ApJ...761...72M,2012ApJ...749...98T,2013arXiv1301.1326L} leading to the expectation of violent fluid motions and turbulence in the mantle as the shock is revived and moves outwards. Like collective and shock effects, turbulence is another supernova feature that can leave its fingerprints upon the neutrino burst
\cite{Loreti:1995ae,Fogli:2006xy,Friedland:2006ta,2010PhRvD..82l3004K,2011PhRvD..84h5023R}. 
At first glance turbulence is just a case of a more complicated MSW effect but, upon further reflection, one realizes that the randomness of the profiles means 
the transition probabilities for a particular neutrino - the set of probabilities that relates
the initial state to the state after passing through the supernova - along a given ray are not unique: they will depend upon the exact turbulence pattern seen 
by the neutrino as it travelled through the supernova. The transition probabilities are drawn from a distribution whose properties will 
depend upon the stage of the explosion, the character of the turbulence, and the neutrino energy and mixing parameters. 
When the mixing angle $\theta_{13}$ was unknown it was difficult to make robust statements about the effect of turbulence because at one value of $\theta_{13}$ 
the effects would be negligible, at another the turbulence would be endemic. The recent measurements of 
the last mixing angle $\theta_{13}$ by T2K \cite{2011PhRvL.107d1801A}, Double Chooz \cite{2012PhRvL.108m1801A}, RENO \cite{2012PhRvL.108s1802A} and Daya Bay
\cite{2012PhRvL.108q1803A} are all in the region of $\theta_{13} \approx 9^{\circ}$, significantly higher than the Dighe \& Smirnov \cite{Dighe:1999bi} threshold, and it is now possible to be 
more definitive about the consequences of turbulence. 

In this paper we consider the implications of the recent measurement of $\theta_{13}$ upon the neutrino transition probabilities 
as a function of the turbulence amplitude. Our calculations expand upon the work of Kneller \& Volpe \cite{2010PhRvD..82l3004K} 
upon which we shall rely heavily for the techniques used to calculate the turbulence effects and as reference for our results.
We first describe the calculations we undertook then present our results for the turbulence effects when the turbulence amplitude is small, less than 1\% comparing large and small $\theta_{13}$. 
We then turn to large amplitude turbulence and compute the expectation values of the transition probability distributions in both neutrinos 
and antineutrinos again comparing large and small $\theta_{13}$. We finish with a summary and our conclusions.


\section{Description of the calculations}

The quantities we are interested in calculating are the probabilities that some initial neutrino state $\ket{\nu(x)}$ at $x$ is later detected as the
state $\ket{\nu(x')}$ at $x'$. These probabilities are computed from the $S$-matrix which relates the initial and final states via $\ket{\nu(x')} =
S(x',x)\,\ket{\nu(x)}$. The $S$-matrix is found by solving the equation
\begin{equation}
\imath \frac{dS}{dx} = H\,S
\end{equation} 
where $H$ is the Hamiltonian. In matter the Hamiltonian is composed of at least two terms: the vacuum contribution $H_0$ and the MSW potential $V$.
When solving for $S$ one must work in a particular basis and the basis determines the structure of the terms in the Hamiltonian.
In the `mass' basis the vacuum Hamiltonian is diagonal and described by two mass squared differences $\delta m_{ij}^2 = m_i^2 - m_j^2$ and the neutrino energy
$E$. Through this paper we shall use the values of $\delta m_{21}^2 = 8 \times 10^{-5}\;{\rm eV^2}$, $|\delta m_{32}^2| = 3 \times 10^{-3}\;{\rm eV^2}$, $\sin^{2}
2\theta_{12}=0.83$ and $\sin^{2} 2\theta_{23}=1$ which are consistent with present experimental values. In the flavor basis the off-diagonal elements are non-zero leading to the phenomenon of flavor
oscillations. The two bases are related by the Maki-Nakagawa-Sakata-Pontecorvo \cite{Maki:1962mu,Nakamura:2010zzi} unitary matrix parametrized by three mixing
angles, $\theta_{12}$, $\theta_{13}$ and $\theta_{23}$, a CP phase and two Majoranna phases.    

In contrast, the MSW potential is diagonal in the flavor basis because the matter picks out the neutrino flavors. The common neutral current contribution
to the MSW potential may be dropped because it leads only to a global phase which is unobservable leaving just the charged current potential, $\sqrt{2} G_F
n_e(r)$ where $G_F$ is the Fermi constant and $n_e(r)$ the electron density, which affects just the electron neutrino/antineutrino i.e. the element $V_{ee}$. 
In addition to the MSW potential, it has been found that the neutrino density in
supernovae is so high that an additional potential due to neutrino self-interactions must be included. This neutrino self-coupling has been shown
to lead to very interesting behavior but for our purposes the self-interaction is negligible when the turbulent region in
the star has moved beyond $1000\;{\rm km}$ so we shall ignore this contribution. 

When the vacuum and matter terms are added together the Hamiltonian is neither diagonal in the mass nor the flavor bases so 
one would expect oscillations of both the flavor and mass probabilities. These oscillations are a source of potential confusion for any analysis.
A basis can be found which diagonalizes $H$ for a given value of the electron density in the sense that there is a matrix $U$ such that 
$U^{\dagger} H U = K$ where $K$ is the diagonal matrix of eigenvalues. This basis is known as the matter basis which becomes the mass basis (up to arbitrary
phases) when the MSW potential disappears. The matter mixing matrix $U$ which achieves this diagonalization depends upon the position through the star
therefore $dU/dx \neq 0$ in general. The non-zero derivative of the matter mixing matrix re-introduces off-diagonal elements into the matter basis
Hamiltonian which will lead to mixing between the matter basis states if they become large. We refer the reader to Kneller \& McLaughlin
\cite{2009PhRvD..80e3002K} and Galais, Kneller \& Volpe \cite{2012JPhG...39c5201G} for a more detailed description 
of the matter mixing matrix. We shall report our results using the matter basis states throughout this paper. 

\begin{figure}
\includegraphics[clip,width=\linewidth]{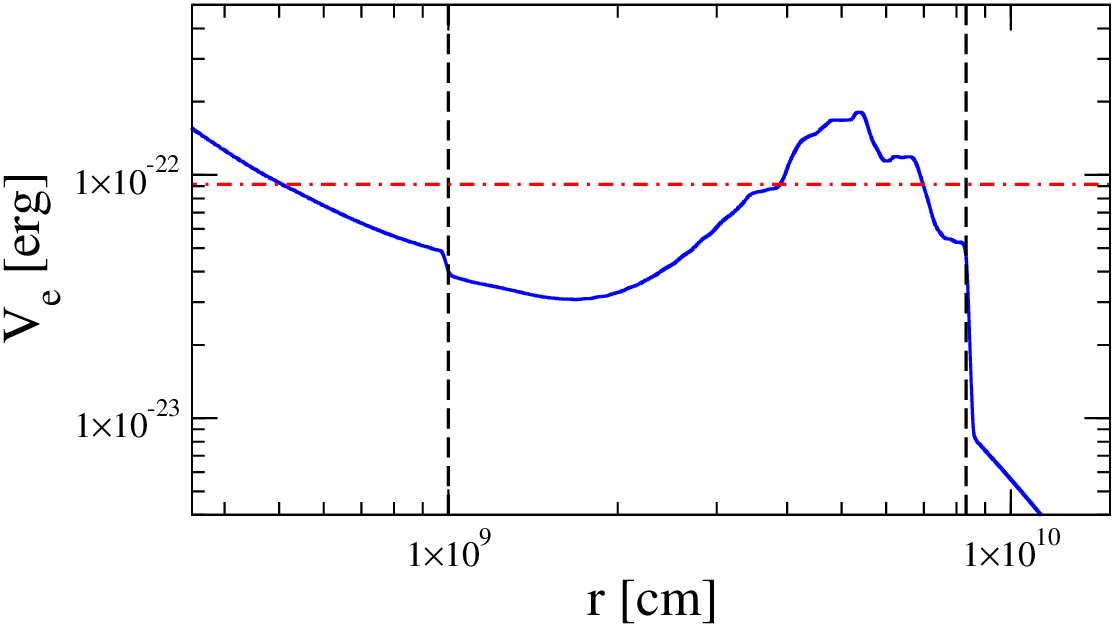}
\caption{The MSW potential as a function of distance through a supernova taken from a hydrodynamical simulation. The vertical lines indicate the positions of
the reverse and froward shock in the profile. The horizontal dashed-dotted line is the two-flavor resonance density for a $25\;{\rm MeV}$ neutrino with mixing angle
$\sin^2 2\theta = 0.1$ and mass splitting $\delta m^{2} = 3\times 10^{-3} \;{\rm eV^2}$ \label{fig:profile}}
\end{figure}
Next we must introduce the turbulent density profile though which the neutrinos will propagate. 
Ideally one would like to use density profiles taken from
multi-dimensional simulations but at the present time that is not possible. The current multi-dimensional simulations do not extend out to the region of $r \gtrsim 10^{4}\;{\rm km}$ where the turbulence would have its greatest effects because the matter there has little bearing upon the explosion, and even if they did, they do not run to sufficiently late post-bounce times to see the shock move out there. Finally, the dynamic scale the simulations would
need to cover would be of order forty to fifty decibels - four to five orders of magnitude - because the neutrino oscillation wavelength is significantly smaller than the radius in the
high-density resonance region and beyond. For these reasons the effect of the turbulence upon the neutrinos is most often modelled as a random field. 
We adopt a one-dimensional supernova profile from a hydrodynamical simulation and in order to facilitate comparison
the profile we select is taken from Kneller, McLaughlin \& Brockman and is the same profile used in Kneller \& Volpe. 
This profile is shown in figure (\ref{fig:profile}). In the figure we find two shocks: the forward shock at $r_s$ formed from the core bounce, and the reverse
shock at $r_r$ formed by the wind created above the proto-neutron star running into the material ahead of it. In multi-dimensional simulations of supernova
both these shock fronts are distorted leading to strong turbulence in the region between them. But for this paper we shall use neutrino energies and mixing
parameters such that the H resonance density does not intersect the shocks. The reason we avoid the shocks is twofold. Hydrodynamical simulations typically
yield `soft' shocks that do not cause transitions between the neutrino states if the mixing angle is too big. This lack of a transition is unphysical. The second reason is that we wish to focus solely upon
the turbulence effect and diabatic MSW transitions caused by the shocks complicates the interpretation. For these reasons we will use $25\;{\rm MeV}$
for the neutrino energy and the two-flavor resonance density for a $25\;{\rm MeV}$ is shown in the figure. The reader can verify that it does not intersect
the profile at either shock.
\begin{figure}
\includegraphics[clip,width=\linewidth]{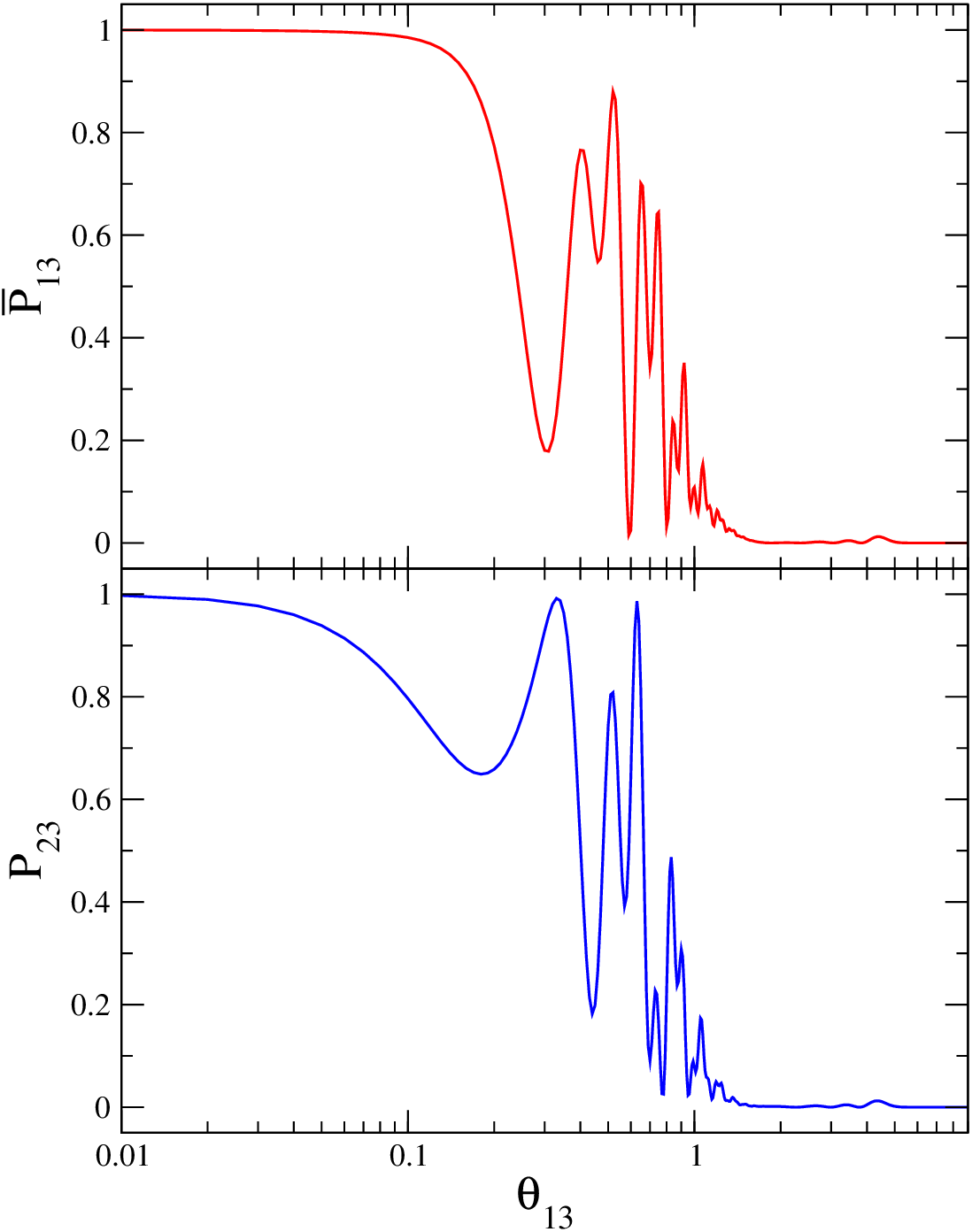}
\caption{The transition probability $\bar{P}_{13}$ in an inverted hierarchy (top panel) and $P_{23}$ in a normal hierarchy (lower panel) after passing through the
density profile shown in figure (\ref{fig:profile}) for a $25\;{\rm MeV}$ neutrino as a function of the mixing angle $\theta_{13}$  \label{fig:P23,P13barvstheta13}}
\end{figure}
For future reference, the transition probability $P_{23}$ for a normal hierarchy and $\bar{P}_{13}$ for an inverted hierarchy are very close to zero for this
neutrino energy, and mass splitting of $\delta m^{2}= 3\times 10^{-3}$ and a mixing angle given by $\sin^{2}2\theta_{13}=0.1$. The reader may be surprised to
see that the figure indicates the transition probabilities change from the diabatic limit, $\bar{P}_{13}=1$ or $P_{23}=1$, to the adiabatic limit
$\bar{P}_{13}=0$ or $P_{23}=0$ is not monotonic in the H resonance channel because of the aforementioned presence of the multiple H resonances in the
profile. The multiple resonances leads to an interference effect which is sensitive to $\theta_{13}$ when $0.1^{\circ} \lesssim \theta_{13} \lesssim 
1^{\circ}$ for this neutrino energy, profile and mass splitting \cite{Kneller:2005hf,Dasgupta:2005wn}.

\subsection{Modelling the turbulence}

\begin{figure*}
\includegraphics[clip,width=0.45\linewidth]{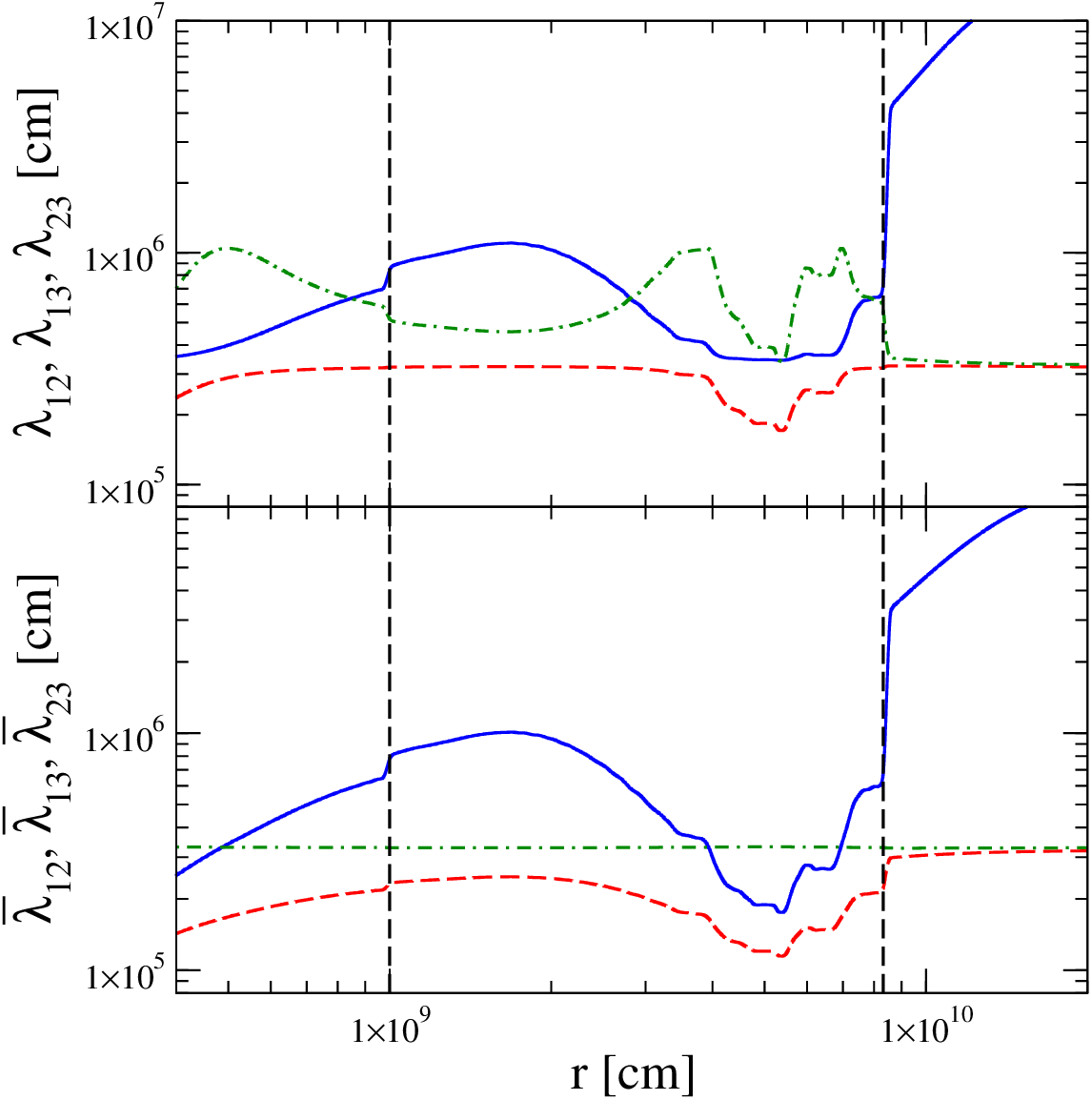}
\includegraphics[clip,width=0.45\linewidth]{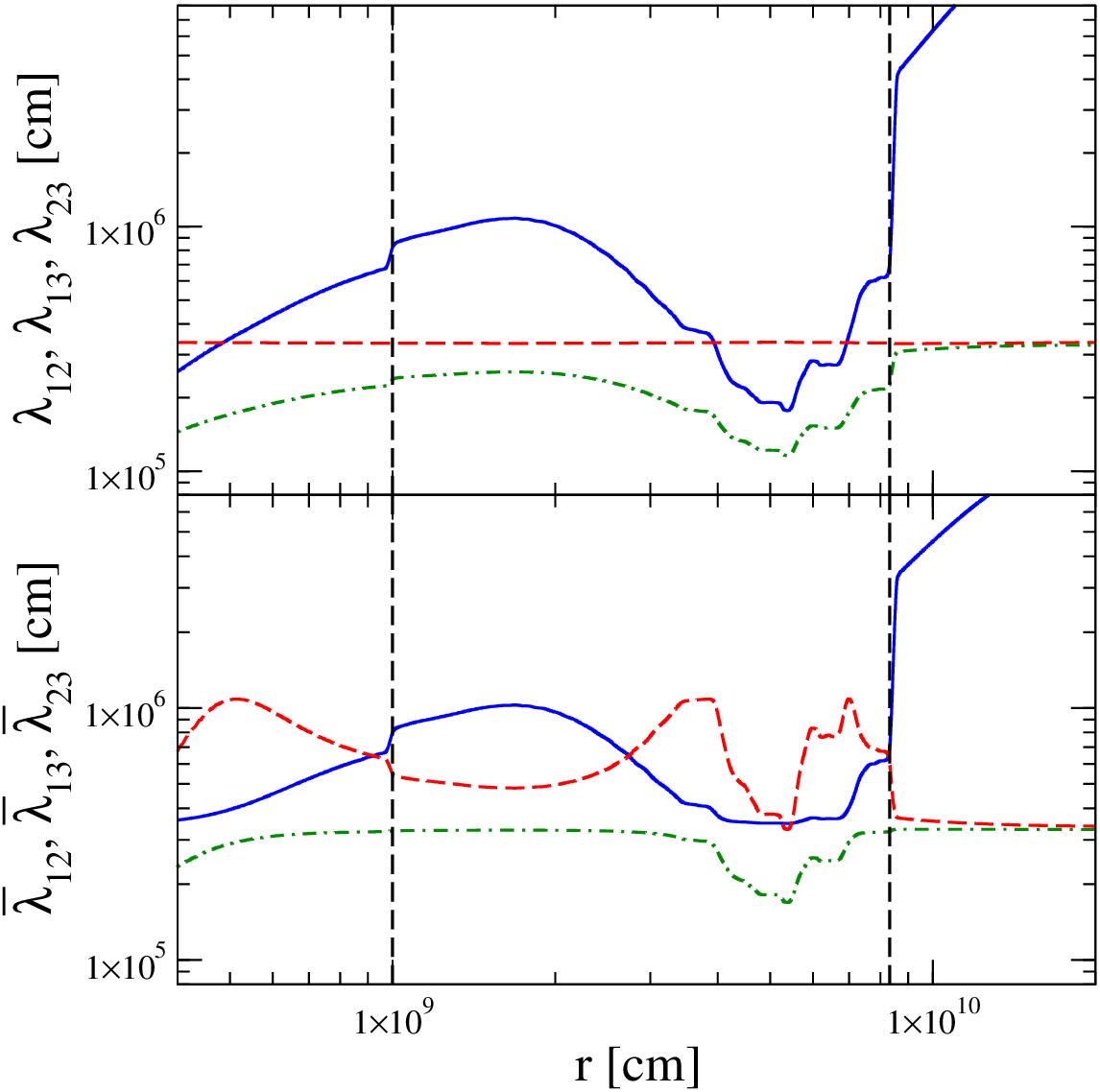}
\caption{The reduced wavelength of the splitting between the eigenvalues. The density profile is that shown in figure (\ref{fig:profile}), the 
neutrino energy is $25\;{\rm MeV}$ and the neutrino mixing angle $\theta_{13}$ is given by $\sin^{2} 2\theta_{13}=0.1$. In the left panel we show the normal hierarchy
case, in the right panel the inverted and the top row of each is for the neutrinos and the bottom row for antineutrinos. \label{fig:deltak}}
\end{figure*}
The turbulence is introduced by multiplying the profile in the region between the reverse and forward shocks by a factor $1+ F(r)$ where $F(r)$ is a
Gaussian random field with zero mean. Since the quality of our results in this entire paper rests firmly upon us doing this well, it is worth
our effort to explain carefully how $F(r)$ was constructed.
The random field is represented using a Fourier series i.e.~. 
\begin{equation}\label{eq:F3D}
\begin{split}
F(r)&=C_{\star}\,\tanh\left(\frac{r-r_r}{\lambda}\right)\,\tanh\left(\frac{r_s-r}{\lambda}\right) \\
& \times\sum_{n=1}^{N_k}\,\sqrt{V_{n}}\left\{ A_{n} \cos\left(k_{n}\,r\right) + B_{n} \sin\left(k_{n}\,r\right) \right\}. 
\end{split}
\end{equation}
for radii between $r_r \leq r \leq r_s$ and zero outside this range. The two radii $r_r$ and $r_s$ are the positions of the reverse and forward shock respectively
found in the underlying profile. In this equation the parameter $C_{\star}$ sets the amplitude of the fluctuations. The two $\tanh$ 
terms are included to suppress fluctuations close to the shocks and prevent discontinuities at $r_s$ and $r_r$, and the parameter $\lambda$ is a
scale over which the fluctuations reach their extent size. We set $\lambda=100\;{\rm km}$. In the second half of equation
(\ref{eq:F3D}) the members of the set of co-efficients $\{A\}$ and $\{B\}$ are independent standard Gaussian random variates with zero mean thus ensuring the
vanishing expectation value of $F$. The $N_{k}$ wavenumbers form a set $k_{n}$ with power spectrum $E(k)$ and, finally, the parameters
$V_{n}$ are k-space volume co-efficients to which we return shortly. 
The power spectrum of the random field was selected to be 
\begin{equation}
E(k) = \frac{(\alpha-1)}{2\,k_{\star}} \left( \frac{k_{\star}}{|k|}\right)^{\alpha}\, \Theta(|k|-k_{\star}). \label{eq:E1D}
\end{equation}
Here $k_{\star}$ is the cutoff scale, $\alpha$ is the spectral index and $\Theta$ is the Heaviside step function.
Throughtout this paper we shall use a wavenumber cutoff $k_{\star}$ set to $k_{\star}=\pi/(r_s-r_r)$ i.e.\ a wavelength twice the distance between the
shocks and we shall adopt the Kolmogorov spectrum where $\alpha=5/3$. 
The method of fixing the $N_k$ $k$'s, $V$'s, $A$'s and $B$'s for a realization of $F$ is `variant C' of the Randomization Method
described in Kramer, Kurbanmuradov, \& Sabelfeld \cite{2007JCoPh.226..897K}.
This Randomization Method partitions the k-space into $N_{k}$ regions and from each we select a random wavevector using the power-spectrum, $E(k)$, as a
probability distribution. The volume parameters $V_{n}$ are the integrals of the power spectrum over each partition if the power spectrum is normalized to unity.
Variant C of the Randomization Method divides the k-space so that the number of partitions per decade is uniform over $N_d$ decades starting from a cutoff scale $k_{\star}$.
The logarithmic distribution of the modes is designed to ensure the quality of the agreement between the exact statistical behavior of the field and
that of an ensemble of realizations is uniform over a the range of lengthscales considered i.e.~ it is scale invariant. This feature is
important for our study because the oscillation wavelength of the neutrinos is constantly changing as the density evolves. 
The evolution of the reduced oscillation wavelengths for the neutrinos - $\lambda_{ij} =1/|\delta k_{ij}|$  - and antineutrinos - $\bar{\lambda}_{ij}
=1/|\delta \bar{k}_{ij}|$ - where $\delta k_{ij}$ and $\delta\bar{k}_{ij}$ are the differences between the eigenvalues $i$ and $j$ of the neutrinos and
antineutrinos respectively - as a function of distance through the profile are
shown in figure (\ref{fig:deltak}) for both a normal and an inverse hierarchy when the mixing angle $\theta_{13}$ is set to $\sin^{2} 2\theta_{13}=0.1$
and the energy is $E=25\;{\rm MeV}$. Again the reverse and forward shocks are indicated by the two vertical dashed lines. This figure can be used to determine
a suitable value for $N_d$ because we observe that in the region between the shocks the typical wavelengths are $\gtrsim 1\;{\rm km}$ which is the 
minimum lengthscale we need to cover \cite{Friedland:2006ta,2012arXiv1202.0776K}. This is approximately
4 orders of magnitude smaller than the turbulence cut-off scale $1/k_{\star}$ thus we deduce that we need to pick $N_d \geq 4$ to cover the necessary decades in k-space.

With $N_d$ determined we now seek a suitable value of $N_k$ by requiring that the statistical properties of an ensemble of random field realizations closely
match the exact properties for the field. The statistical property we compute is the second
order structure function $G_2(\delta r )$ given by 
\begin{equation}
G_2(\delta r) = \langle F(r + \delta r) - F(r) \rangle^2 \label{eq:G2}
\end{equation}
where $\delta r$ is the separation between two radial points. The function $G_2(\delta r)$ is related to the two-point correlation function $B(\delta
r)$ via $G_2(\delta r)/2 = 1-B(\delta r)$ and for the power spectrum we have adopted we can compute the two-point correlation function analytically to be 
\begin{widetext}
\begin{equation}
B(\delta r) = \frac{(\alpha-1)}{2}\,\left(2\pi\,k_{\star}\,\delta r \right)^{\alpha-1} \left\{\exp\left(\frac{\imath\pi\alpha}{2}\right)\,\Gamma(1-\alpha,2\imath\pi\,k_{\star}\,\delta r) +
\exp\left(\frac{\imath\pi\alpha}{2}\right)\,\Gamma(1-\alpha,-2\imath\pi\,k_{\star}\,\delta r) \right\}. \label{eq:zcorrelation}
\end{equation}
\end{widetext}  
where $\Gamma(n,x)$ is the incomplete Gamma function.
\begin{figure}[t]
\includegraphics[clip,width=\linewidth]{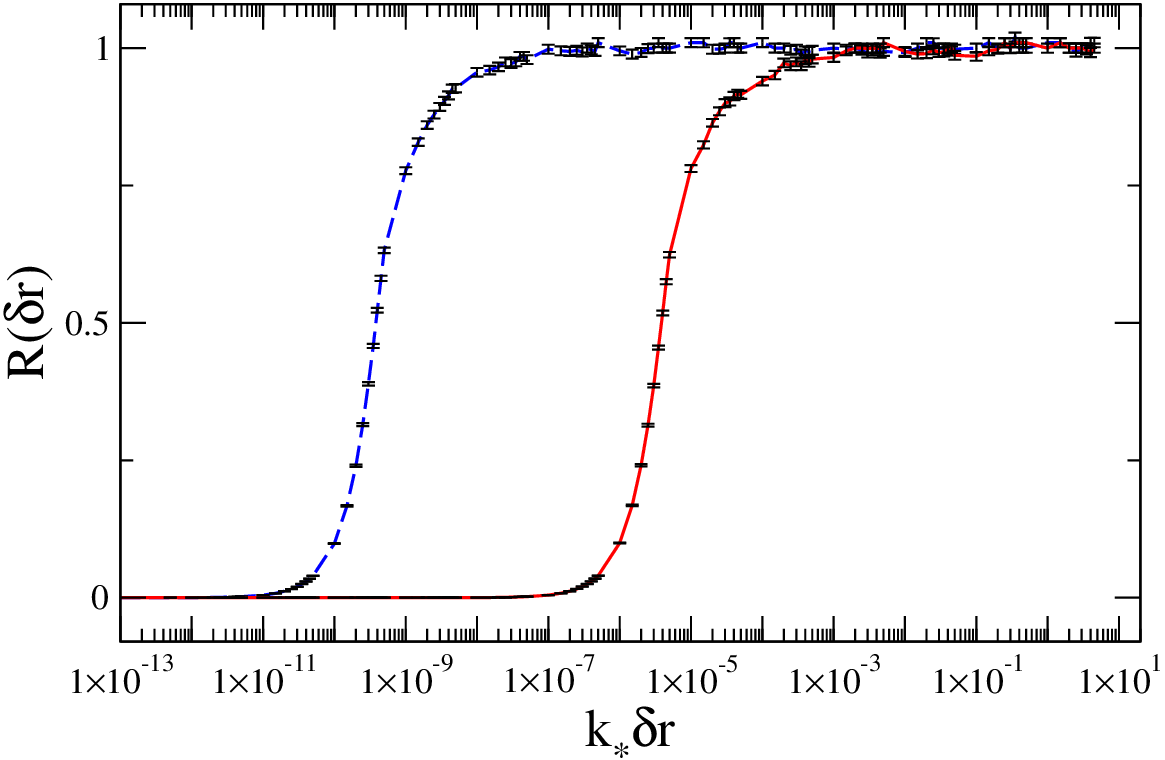}
\caption{The ratio $R(\delta r)$ of the numerically calculated structure function to the analytic result as a function of $k_{\star} \delta r$. 
The two curves in the figure correspond to $\{N_k,N_d\}=\{50,5\}$ (red solid) and $\{N_k,N_d\}=\{90,9\}$ (blue dashed).
At every $k_{\star} \delta r$ we generated $30,000$ realization of the field and the error bar on each point is the standard deviation of the mean. \label{fig:GRF1D}} 
\end{figure}
In figure (\ref{fig:GRF1D}) we show the ratio $R(\delta r)$ of the numerically calculated structure function to the exact solution as a function of
the scale $k_{\star}\delta r$ when we use either $N_{k}=50$ wavenumbers spread over $N_d=5$ decades or $N_k=90$ wavenumbers over $N_d=9$ decades. The numerical
calculation is the average of $30,000$ realizations of the turbulence and the error bar on each point is the standard deviation of the sample mean. The figure indicates
that the method we use to generate random field realizations reproduces the analytic results for the structure function very well and with high efficiency
because good agreement between the statistics of the ensemble and the exact result requires just $N_{k}/N_d = 10$. In fact, like Kramer, Kurbanmuradov, \&
Sabelfeld \cite{2007JCoPh.226..897K} before us, we find even $N_{k}/N_d$ ratios of just $N_{k}/N_d \sim 2-3$ are sufficient to give acceptable agreement. We
re-assure the reader we shall stick with $N_{k}/N_d = 10$.


\section{Results}

With the construction of the random fields in place we can proceed to generate a turbulent profile and propagate neutrinos and antineutrinos through it. 
This construction and propagation recipe is then repeated a minimum of one thousand times - sometimes much larger - to construct an ensemble of transition
probabilities of size $N$. Once we have our sample we can then go ahead and compute means $\langle P_{ij}\rangle$, variances $\sigma_{ij}$, etc.
The hierarchy will be set to normal and we shall comment on how our results translate to the inverted hierarchy. The neutrino energy will be fixed at
$E=25\;{\rm MeV}$, typical of supernova neutrino energies. The turbulence effects - or lack of them - when using a value of $\theta_{13}$ close to the present measurements was not fully explored 
in previous studies so to make a connection with previous works, and to explain why a large value of
$\theta_{13}$ gives the results that it does, we shall consider multiple values of $\theta_{13}$ in order to show what other possibilities would have produced in contrast. 


\subsection{Small amplitude turbulence}
\begin{figure}[b]
\includegraphics[clip,width=\linewidth]{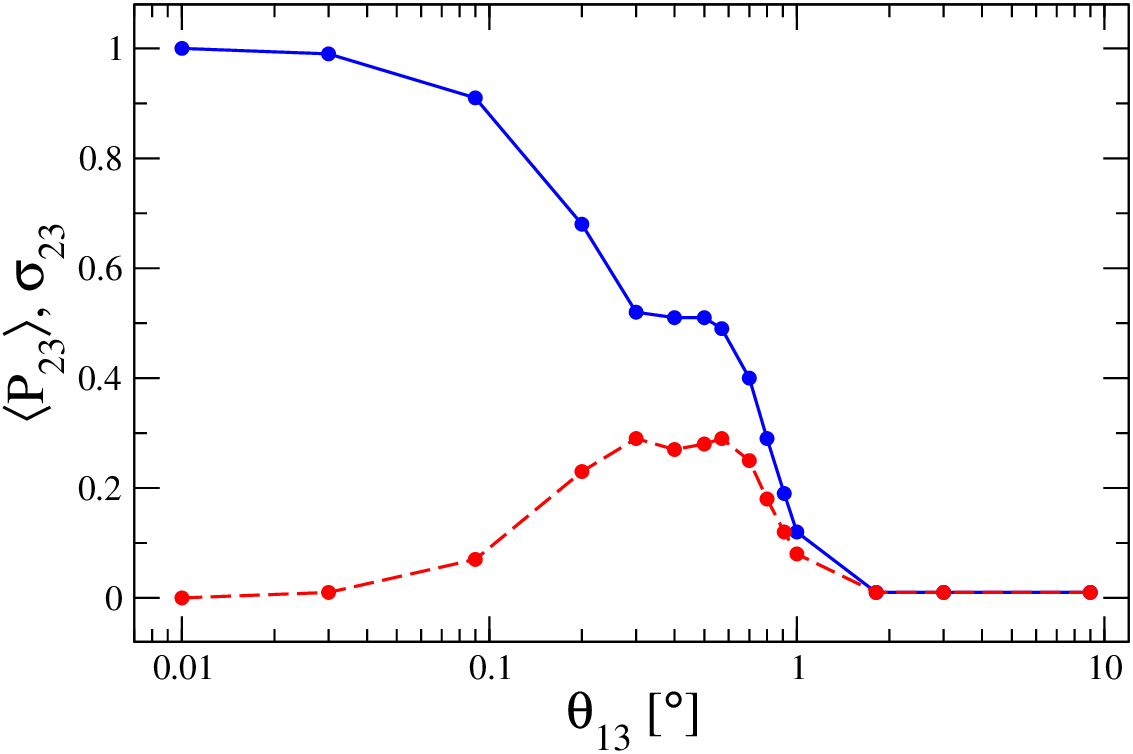}
\caption{The mean of the transition probability, $\langle P_{23} \rangle$ (solid line) and the standard deviation $\sigma_{23}$ (dashed line) 
as function of the mixing angle $\theta_{13}$. The turbulence amplitude is set to $C_{\star}=1\%$ and $\{N_k,N_d\}=\{50,5\}$
and the neutrino energy is $25\;{\rm MeV}$. \label{fig:P23vstheta13}}
\end{figure}
For small amplitude turbulence only the H-resonant channel is affected: mixing between states $\nu_2$ and $\nu_3$ for a normal hierarchy and states
$\bar{\nu}_1$ and $\bar{\nu}_3$ for an inverted. Previously it has been found that effects could appear in the neutrinos even for turbulence amplitudes in the range $10^{-5} \lesssim C_{\star} \lesssim 0.1$
when the mixing angle was set at $\sin^2 2\theta_{13} = 4\times 10^{-4}$. If we allow the value of $\theta_{13}$ to float then we find the normal hierarchy H resonance channel transition probability 
$P_{23}$ can become more or less diabatic. This can be explained from the behavior of the diabaticity parameter $\Gamma_{23}$
\cite{2009PhRvD..80e3002K}, which characterizes the degree of mixing in the H resonance channel for a normal hierarchy. This quantity is inversely proportional to the
difference $\delta k_{23}$ between the eigenvalues $k_{2}$ and $k_{3}$ and proportional to the derivative of the matter mixing angle $\tilde{\theta}_{13}$. Increasing $\theta_{13}$
increases the eigenvalue splitting and also makes the resonance `wider' in the sense that the change between the limiting values of
the matter mixing angle $\tilde{\theta}_{13}$ occurs over a greater extent reducing the matter angle derivative. Both effects decrease the diabaticity and, for these
reasons, reaching the depolarization limit for $P_{23}$ becomes more difficult if the domain of turbulence is fixed as is the case here.
If the profile were changed so as to allow a larger turbulence region then eventually one should expect to reach the depolarization limit no matter what
the mixing angle. A similar argument applies when $\theta_{13}$ becomes small: now the diabaticity increases as $\theta_{13}$ decreases because the splitting between the
eigenvalues at the resonance decreases and the transition occurs more rapidly. Either way, as $\theta_{13}$ varies the distributions for the
transition probability $P_{23}$ will differ from the uniform distributions seen in Kneller \& Volpe leading to subsequent evolution of the
expectation values and distribution variances. This evolution with $\theta_{13}$ is seen in figure (\ref{fig:P23vstheta13}) where we plot the mean value $\langle P_{23} \rangle$
and the standard deviation of the samples from a single emission point as a function of $\theta_{13}$. The reader should compare the evolution of $\langle
P_{23} \rangle$ in this figure with that in figure (\ref{fig:P23,P13barvstheta13}). The mean value of $\langle P_{23} \rangle$ has
an inflection region between $0.1^{\circ} \lesssim \theta_{13} \lesssim 1^{\circ}$: for the smaller values of $\theta_{13}$ we see $\langle P_{23}
\rangle  > 1/2$, for the larger $\langle P_{23} \rangle  < 1/2$ and for $\theta_{13} \sim 9^{\circ}$ the mean value of $P_{23}$ is almost zero. One also
observes how the sample standard deviation changes as $\theta_{13}$ varies and sees that it is maximal at $\sigma_{23} = 0.28$ for the range $0.1^{\circ}
\lesssim \theta_{13} \lesssim 1^{\circ}$ and almost zero when $\theta_{13} \sim 9^{\circ}$. This figure shows how the measurement of $\theta_{13}$ has brought 
clarity to the issue of turbulence and supernova neutrinos. For $\theta_{13}$ outside the range $0.1^{\circ} \lesssim \theta_{13} \lesssim 1^{\circ}$ the 
distribution of $P_{23}$ is essentially a delta function at either zero or unity; for $\theta_{13}$ inside the range $0.1^{\circ} \lesssim \theta_{13} \lesssim 1^{\circ}$
the distribution is uniform. Thus when $\theta_{13}$ was unknown it was impossible to determine whether the effect of small amplitude turbulence was negligible or overwhelming.
The measurement of a large value of $\theta_{13}$ indicates it is the former and the result has consequences for the observability of spectral features in the 
next Galactic supernova burst signal.    


\subsection{Large Amplitudes}

\subsubsection{The neutrino mixing channels}
For large amplitudes, $C_{\star} \gtrsim 0.1$, the effects of turbulence are no longer restricted to the H resonance channel but appear in numerous places. The
first effect worth noting is that the distribution of the H-resonance channel transition probability, $P_{23}$ in the case of a normal hierarchy, becomes
independent of $\theta_{13}$. This can be seen in figure (\ref{fig:P12,P13,P23vsCstar}) where the reader will observe the evolution of
the mean value of this transition probability $\langle P_{23} \rangle$ as a function of $C_{\star}$. For the two values of $\theta_{13}$ considered, the 
spread in $\langle P_{23} \rangle$ at small amplitudes has disappeared by $C_{\star} \sim 0.3$. 
We also notice that around this same turbulence amplitude there begins the shift to three-flavor depolarization where $\langle P_{23}\rangle = 1/3$. 

In addition to the changes in the H-resonance channel we also begin to observe mixing in the L-resonance channel, between $\nu_1$ and $\nu_2$ as 
the amplitude grows. This simultaneous mixing between $\nu_1$ and $\nu_2$ and $\nu_2$ and $\nu_3$ breaks HL factorization and Kneller \& Volpe gave two
examples were given that explicitly showed broken HL factorization. For the neutrino mixing parameters we are using, the ratio of H and L resonance
densities (using the two-flavor formula) is $\rho_H/\rho_L = (\delta m_{23}^{2} \cos 2\theta_{13}) / (\delta m_{12}^{2} \cos 2\theta_{12}) \approx 90$. This
large ratio would seem to imply that we need fluctuations of order $F \sim 1$ because only if $F=-0.99$ would the density fluctuation give $\rho_{H}(1+F)
\approx \rho_L$. Three effects soften this requirement: the L resonance has a large width - $\Delta \rho_L /\rho_L = \tan 2\theta_{12} \sim 1$, the density in
the turbulence region can be much lower than the resonance density $\rho_{H}$ for the given neutrino energy - see figure (\ref{fig:profile}), and, finally,
our choice of a Gaussian random field for the turbulence will ensure that large fluctuations will occur occasionally no matter what we use 
for the amplitude, larger amplitudes just make the extremal fluctuations more probable.
\begin{figure}
\includegraphics[clip,width=\linewidth]{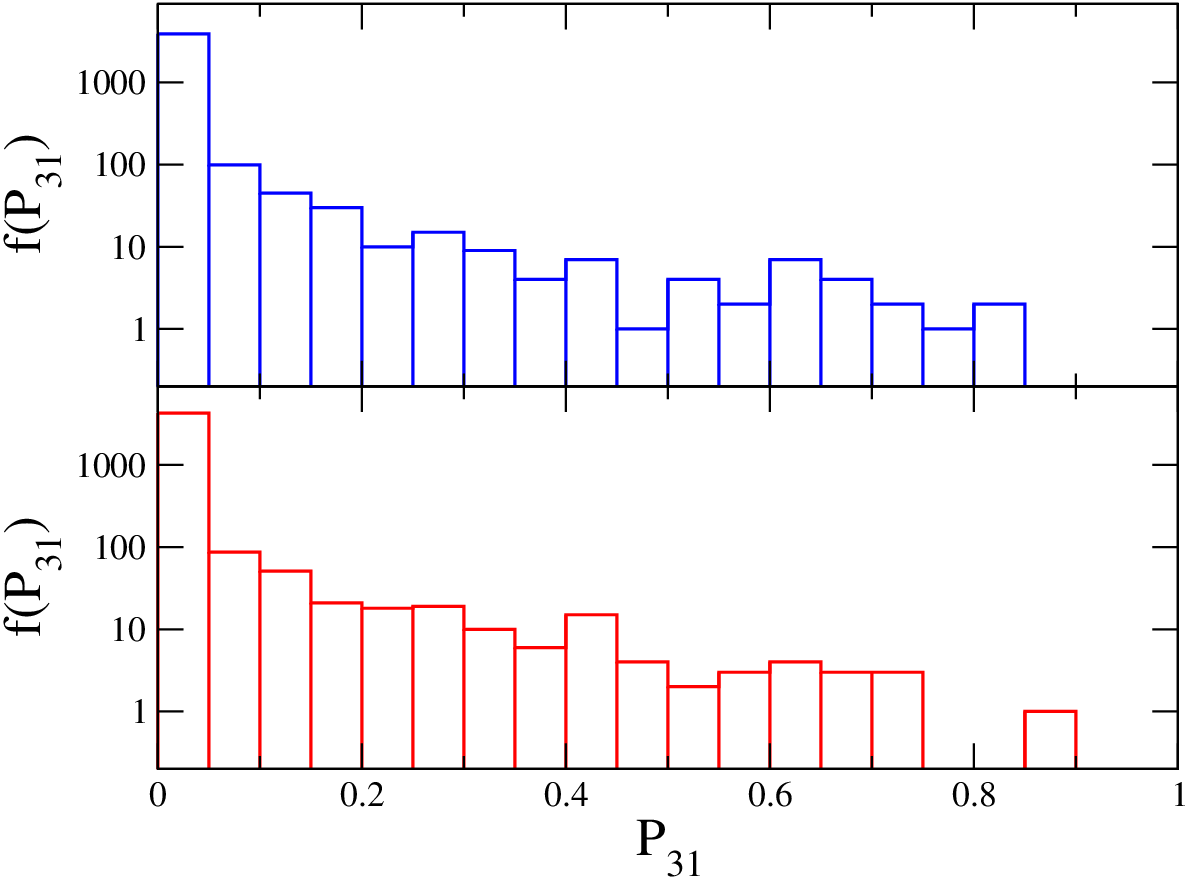}
\caption{The frequency distribution of the transition probability $P_{31}$ for a turbulence amplitude is set to $C_{\star}=0.3$. The neutrino energy is $25\;{\rm MeV}$ 
and we used $N_{k} = 50$, $N_d=5$ for the turbulence generator. The top panel used a
value of $\theta_{13}$ given by $\sin^2 2\theta_{13}=0.1$ and the bottom panel is for $\sin^2 2\theta_{13}=4\times 10^{-4}$. \label{fig:f(P31)} }
\end{figure}
The clearest signature of broken HL factorization is a non-zero transition probability $P_{31}$ because only if HL factorization is broken can we generate
an effective mixing between $\nu_1$ and $\nu_3$. To see this we consider the $S$-matrices for the case of factored HL resonances
and broken factorization. The S-matrix for passing through one or several H resonances, $S_H$, has the genreal form
\begin{equation}
 S_H = \left(\begin{array}{ccc} 1 & 0 & 0 \\ 0 & \alpha_H & \beta_H \\ 0 & -\beta_H^{\star} & \alpha_H^{\star} \end{array} \right)  
\end{equation}  
where $\alpha_H$ and $\beta_H$ are Cayley-Klein parameters. Similarly the S-matrix for L resonances, $S_L$, is 
\begin{equation}
 S_L = \left(\begin{array}{ccc} \alpha_L & \beta_L & 0 \\ -\beta_L^{\star} & \alpha_L^{\star} & 0 \\ 0 & 0 & 1\end{array} \right)  
\end{equation}  
where $\alpha_H$ and $\beta_H$ are Cayley-Klein parameters for the L resonance. If HL factorization holds then the $S$-matrix which describes the evolution for
the neutrino through the entire profile is $S = S_L S_H$. If all the L resonances occur after the H resonances then we find the transition probability $P_{31} = |S_{31}|^2$ is
identically zero. But if additional H and L resonances occur, denoted by $S'_H$ and $S'_L$, then the $S$-matrix describing the neutrino evolution is of the form $S = S'_L S'_H
S_L S_H$ and $P_{31}=|\beta_{H}'|^{2}|\beta_{L}|^{2}$ will be non-zero. In figure (\ref{fig:f(P31)}) we show frequency distributions of the transition
probability $P_{31}$ for neutrinos at two values of $\theta_{13}$ when $C_{\star}=0.3$. The distributions are clearly non-zero
for non-zero $P_{31}$ as expected if HL factorization were broken. The distributions fall rapidly as something like inverse-power laws, $f(P_{31})
\propto 1/P_{31}^n$ with $n\sim 2$ or exponentials for this particular calculation.

Other mixing channels which were previously delta-distributed for small turbulence amplitudes - such as $P_{12}$ and $P_{13}$ - also begin to possess similar 
inverse power-law/exponential distributions when $C_{\star} \gtrsim 0.1$ and their means increase quadratically with $C_{\star}$. The evolution of these two
transition probabilities is also shown in figure (\ref{fig:P12,P13,P23vsCstar}). The figure shows that $\langle P_{12} \rangle$ at some fixed $C_{\star}$
increases as $\theta_{13}$ increases but $\langle P_{13} \rangle$ decreases as $\theta_{13}$ increases though, in both cases, the change is not very large. The
anticorrelation between $\langle P_{12} \rangle$ and $\langle P_{13} \rangle$ is a reflection of the unitarity requirement that $\Sigma_j P_{ij} =1$ for a given $i$.

\begin{figure}[t]
\includegraphics[clip,width=\linewidth]{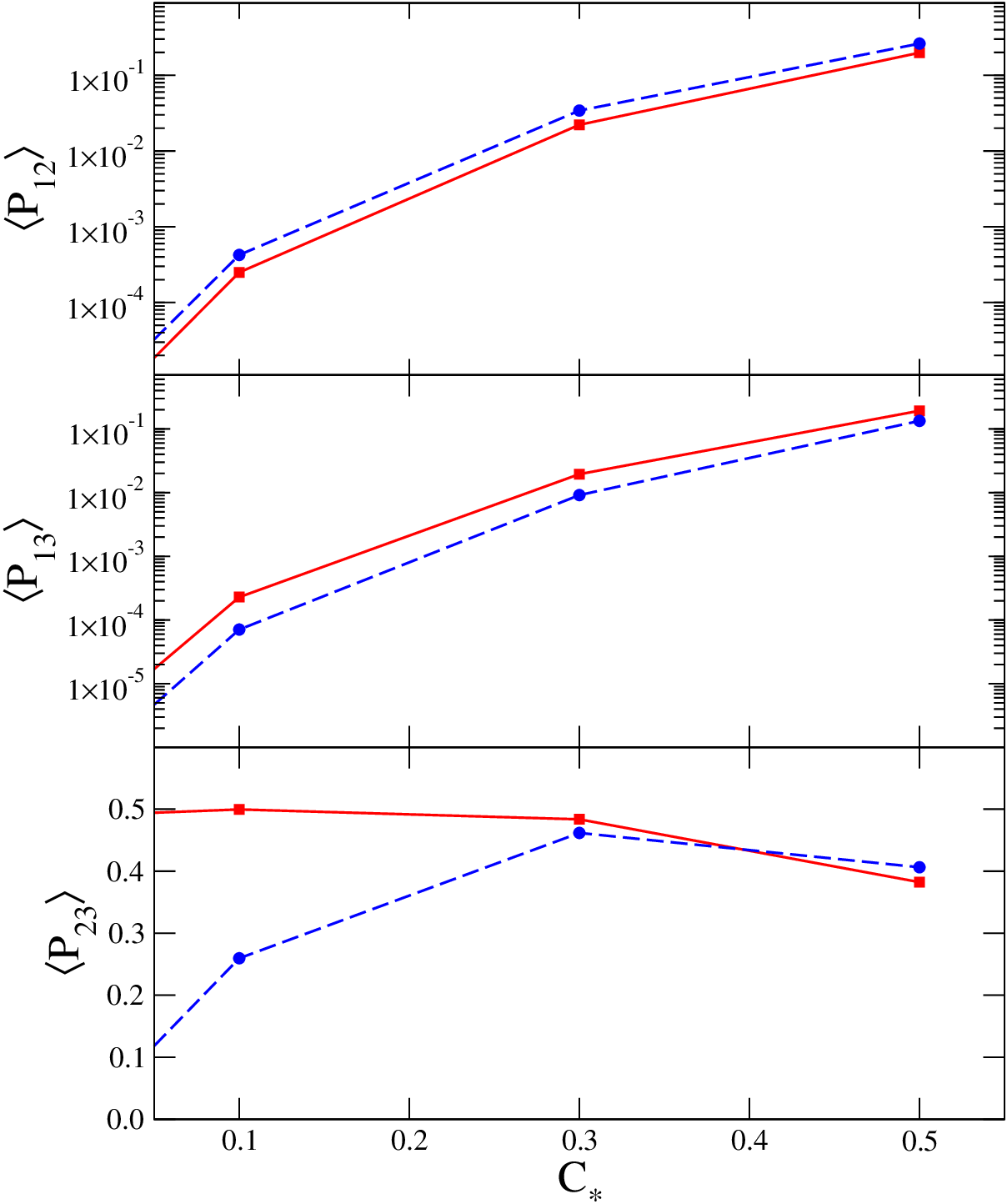}
\caption{The mean of the transition probabilities $P_{12}$ - top panel - $P_{13}$ - center panel - and $P_{23}$ - bottom panel - through anisotropic
turbulence as a function of $C_{\star}$ for antineutrinos emitted from a single point. Each curve corresponds to a different value of $\theta_{13}$: $\sin^2
2\theta_{13}=4\times 10^{-4}$ are squares joined by solid lines, $\sin^2 2\theta_{13}=0.1$ are circles joined by dashed lines. \label{fig:P12,P13,P23vsCstar} }
\end{figure}


\subsubsection{The antineutrino mixing channels}

\begin{figure}
\includegraphics[clip,width=\linewidth]{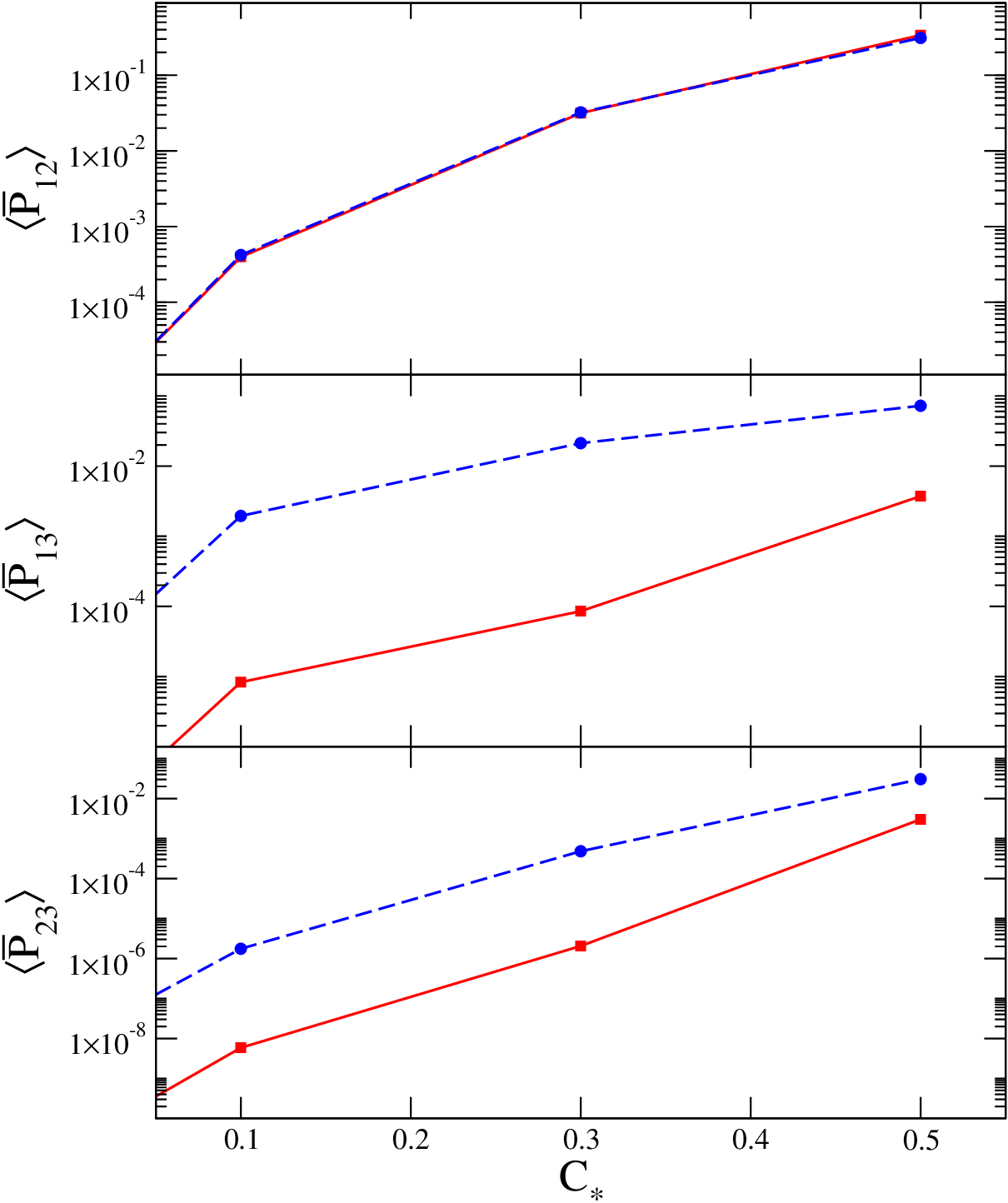}
\caption{The mean of the transition probability $\bar{P}_{12}$ - top panel - $\bar{P}_{13}$ - center panel - and $\bar{P}_{23}$ - bottom panel - 
as a function of $C_{\star}$ for antineutrinos. Each curve corresponds to a different value of $\theta_{13}$:
$\sin^2 2\theta_{13}=4\times 10^{-4}$ are squares and $\sin^2 2\theta_{13}=0.1$ are circles. \label{fig:aP12,a13,a23vsCstar} }
\end{figure}
In addition to breaking HL factorization, large amplitude turbulence induces effects in the \emph{non-resonance} channels particularly $\bar{\nu}_1
\leftrightarrow \bar{\nu}_2$ regardless of the hierarchy, $\bar{\nu}_1 \leftrightarrow \bar{\nu}_3$ for a normal hierarchy and $P_{23}$ for an inverted
hierarchy. If we stick with considering the normal hierarchy case then we can compute the mean of the non-resonant transition probabilities $\bar{P}_{12}$,
$\bar{P}_{13}$ and $\bar{P}_{23}$ as a function of the mixing angle $\theta_{13}$ and turbulence amplitude $C_{\star}$. These are shown in figure
(\ref{fig:aP12,a13,a23vsCstar}). Like $\langle P_{12}\rangle$ and $\langle P_{13}\rangle$ shown in figure (\ref{fig:P12,P13,P23vsCstar}), the reader will observe that the three
transition probabilities grow rapidly with $C_{\star}$ reaching the levels of $\langle\bar{P}_{12}\rangle \sim \mathcal(10\%)$, $\langle\bar{P}_{13}\rangle
\sim \mathcal(1\%)$ and $\langle\bar{P}_{23}\rangle \sim \mathcal(1\%)$ at $C_{\star} \sim 0.5$. Further comparison with figure
(\ref{fig:P12,P13,P23vsCstar}) reveals $\langle P_{12}\rangle \approx \langle \bar{P}_{12}\rangle$ and the expectation values for the transition probabilities
$\bar{P}_{13}$ and $\bar{P}_{23}$ are both smaller than $\langle P_{13}\rangle$ by roughly an order of magnitude and much more sensitive
to $\theta_{13}$. The expectation value for $P_{13}$ varied by a factor of $\sim 2$ when $\theta_{13}$ allowed to float, here 
$\bar{P}_{13}$ and $\bar{P}_{23}$ change by an $\sim 1-2$ orders of magnitude when increasing from $\sin^2 2\theta_{13}=4 \times 10^{-4}$ to $\sin^2
2\theta_{13}=0.1$. This same sensitivity to $\theta_{13}$ was explained in Kneller \& Volpe as due to the proportionality of the antineutrino
diabaticity parameter $\bar{\Gamma}_{13}$ to the vacuum mixing angle. The current preference for $\theta_{13}$ close to $\theta_{13} \sim 9^{\circ}$
indicates $\langle \bar{P}_{13}\rangle$ and $\langle \bar{P}_{23}\rangle$ can be of order a few percent if $C_{\star} \sim 0.5$.


\section{Summary and Conclusions}

The effects of supernova turbulence upon the flavor composition of neutrinos that pass through it depend upon the numerous parameters that one 
needs to introduce.    
For a neutrino energy of $25\;{\rm MeV}$ and using a supernova density profile taken from a simulation $4.5\;{\rm s}$ post-bounce, turbulence of
amplitude $C_{\star} = 1\%$ only affects the H resonance mixing channel to any appreciable degreee and then only for mixing angles in the range
$0.1^{\circ} \lesssim \theta_{13} \lesssim 1^{\circ}$. For presently favored value of $\theta_{13}$ closer to $9^{\circ}$ there is little effect of small
amplitude turbulence. This result will remain valid for other similar neutrino energies which have their MSW resonances in the region where the turbulence is located. At
later post-bounce times the turbulence region will move outwards and to lower densities affecting the neutrino L resonance. The phenomenology will be similar
to the affect of turbulence upon the H resonance. 
The removal of turbulence effects upon the neutrinos at small amplitudes has important consequences for the prospect of observing signatures of collective and shock wave effects in supernova neutrino burst
signals, which is explored in Lund \& Kneller \cite{KnellerLund}.

For the same neutrino energy and post-bounce epoch, the turbulence effects metastasize as the amplitude increases. In addition, the sensitivity to 
$\theta_{13}$ in the H resonant mixing channel is lost and HL factorization becomes increasingly broken. 
For amplitudes of $C_{\star} \gtrsim 0.3$, and a normal hierarchy the expectation values of the transition probabilities $P_{12}$, $P_{13}$, $P_{23}$ and $\bar{P}_{12}$ are of order 10\% or greter; 
in an inverted hierarchy it is the transition probabilities $P_{12}$, $\bar{P}_{12}$, $\bar{P}_{13}$ and $\bar{P}_{23}$ whose expectation values are of equivalent magnitudes. 
These channels are the most promising for observing the signatures of turbulence.

\acknowledgments
This work was supported by DOE grant DE-SC0006417, the Topical Collaboration in Nuclear Science ``Neutrinos and Nucleosynthesis in Hot and Dense Matter``, DOE grant number DE-SC0004786, 
and an Undergraduate Research Grant from NC State University. 


\end{document}